\documentclass[twocolumn,showpacs,amsmath,amssymb,prd]{revtex4}
\usepackage{epsf}
\usepackage{dcolumn}
\usepackage{bm}
\everymath{\displaystyle}
\begin{document}

\title{High-precision description and new properties of a spin-1 particle in a magnetic field}


\author{Alexander J. Silenko}

\affiliation{Research Institute for Nuclear Problems, Belarusian State University, Minsk 220030, Belarus\\
Bogoliubov Laboratory of Theoretical Physics, Joint Institute for Nuclear Research,
Dubna 141980, Russia}

\date{\today}

\begin{abstract}
The exact Foldy-Wouthuysen Hamiltonian is derived for a pointlike spin-1 particle with a normal magnetic moment in a nonuniform magnetic field. For a uniform magnetic field, it is exactly separated into terms linear and quadratic in spin. New unexpected properties of a particle with an anomalous
magnetic moment are found. Spin projections of particle moving in a uniform magnetic field are not integer and the tensor polarization is asymmetric in the plane orthogonal to the field.
Previously described spin-tensor effects caused by the tensor magnetic polarizability exist not only for nuclei but also for pointlike particles.
\end{abstract}

\maketitle

\section{Introduction}

The problem of a spinning particle in a magnetic field is rather important
for particle physics. A high precision is necessary
not only for a measurement of magnetic moments but also for a search of electric dipole moments
\cite{EDM} and a determination of tensor electric and magnetic polarizabilities \cite{Bar1,Bar3,PRC,PRC2008,PRC2009,dEDMtensor}.
As a result, the existence of exact and high-precision solutions is very important. Almost all known exact results were obtained for a pointlike (structureless) particle in a uniform magnetic field.
Exact solutions were first found in the Dirac representation
for energy levels and wave functions of a relativistic spin-1/2 particle with an anomalous
magnetic moment (AMM) \cite{TBZ} (see also Ref. \cite{Bal} and references therein).
Corresponding solutions were then obtained in the Foldy-Wouthuysen (FW) representation
\cite{JETP1998E,JINRL2008}. This representation restores Schr\"{o}dinger-like forms of Hamiltonians and equations of motion 
and is very convenient for finding a semiclassical approximation and a classical limit of relativistic
quantum mechanics \cite{workFWt,GosPhysLettA,PRA}.

The $g$ factor, $g=2m\mu/(es)$ ($s$ is the spin number), plays an important role. The exact FW transformation has been fulfilled for a Dirac particle ($g=2$) in any \emph{stationary nonuniform}
magnetic field \cite{C} and for a spin-1/2 particle with an AMM moving in the plane orthogonal to a static uniform magnetic field \cite{JETP1998E,JINRL2008}. The exact quantum-mechanical equation of spin motion has also been
derived \cite{JETP1998E}. The exact FW Hamiltonian has been obtained for a spin-1 particle with a normal magnetic moment ($g=2$) in a \emph{uniform} magnetic field \cite{EPJC}.
For a Dirac particle and a spin-1 particle with $g=2$, a conservation of some projections of the polarization operator in a uniform magnetic field has been proven in Refs. \cite{CJP}
and \cite{EPJC}, respectively.

Nevertheless, the above-mentioned results do not exhaust exact FW transformations for a spin-1 particle in uniform and nonuniform magnetic fields.
In the present paper, we obtain new exact and high-precision results. These results allow us to find new polarization properties of spin-1 particles
that can be extended to particles and nuclei with higher spins.

The system of units $\hbar=1,~c=1$ is used. Commutators and anticommutators are denoted by $[\dots,\dots]$ and $\{\dots,\dots\}$, respectively.

\section{Initial Hamiltonian}

An important property of all exact results is their validity for relativistic and ultrarelativistic particles. Such results may be useful for polarized relativistic particle beams in accelerators and storage rings.
Any generalization of exact results obtained for a spin-1 particle in a uniform magnetic field \cite{EPJC} is very important because
magnetic fields in accelerators and storage rings are nonuniform.

A spin-1 particle 
is described by the Proca equations \cite{Pr} added by the Corben-Schwinger term \cite{CS} characterizing the AMM. The electric dipole moment can also be taken into account \cite{Spinunit}.
The Sakata-Taketani transformation \cite{SaTa} brings the initial  equations to the Hamiltonian form. The general Hamiltonian in the
Sakata-Taketani representation has been derived in Ref. \cite{YB}.
For a spin-1 particle in the magnetic field $\bm B$, it is convenient to present this  Hamiltonian in the form \cite{TMPFW}
\begin{equation}  \begin{array}{c} {\cal H}=\rho_3 {\cal M}+{\cal E}+{\cal
O}, ~~~{\cal M}=m+\frac{\bm\pi^2}{2m}-\frac{e\hbar}{m}\bm
S\cdot\bm B, \\
{\cal E}=-\rho_3\frac{e\hbar(g-2)}{2m}\bm S\cdot\bm B, \\
{\cal O}=i\rho_2\left[\frac{\bm\pi^2}{2m}-\frac{(\bm\pi\cdot \bm
S)^2}{m}+\frac{e\hbar(g-2)}{2m}\bm S\cdot\bm B\right].
\end{array} \label{eqXVI} \end{equation}
where $\kappa=g-1={\rm const}$ and $\bm\pi=-i\bm D=-i\nabla-e\bm A$ is the kinetic momentum
operator. Here $\bm S$ is the $3\times3$ spin matrix and $\rho_i~(i=1,2,3)$
are the $2\times2$ Pauli matrices. 
Denotation $\rho_iS_j$ means the
direct product of two matrices. We imply that spin-independent terms are multiplied by the $3\times3$ unit matrix ${\cal I}'$. The even operators ${\cal M}$ and ${\cal E}$ and the odd operator ${\cal O}$ are commutative and anticommutative with the matrix $\beta\equiv\rho_3{\cal I}'$, respectively.
The wave function in the Sakata-Taketani representation is a six-component analog of a Dirac bispinor. For spin-1 particles, the
polarization operator is equal to $\bm\Pi=\rho_3\bm S$. It is
analogous to the corresponding Dirac operator which can be written
in a similar form (see Ref. \cite{ST}): $\bm\Pi=\rho_3\bm\sigma$. When the magnetic field is uniform,
$[{\cal M},{\cal E}]=[{\cal M},{\cal O}]=0$ (the operator ${\cal M}$ is implied to be multiplied by the $2\times2$ unit matrix ${\cal I}$).

\section{Exact results for a spin-1 particle with a normal magnetic moment}

The FW transformation may be exact only for a particle with a normal magnetic moment \cite{EPJC}, when ${\cal E}=0$. The condition of exact FW transformation \cite{PRA} is $[{\cal M},{\cal O}]=0$.
The gyromagnetic ratio $g=2$ is natural for elementary particles of arbitrary spin \cite{ren}.

A charged pointlike spin-1 particle is the $W$ boson.
While the deuteron is a nucleus, its $g$ factor is also close to 2
($g_d = 1.714$). Therefore, a possibility of the exact FW transformation is important for 
deuteron beams in accelerators and storage rings.

When a magnetic field is nonuniform, the operators $\bm\pi$ and $\bm B$ do not commute.
In this case, 
\begin{equation}\begin{array}{c}
[{\cal M},{\cal O}]=-\rho_2\frac{e}{4m^2}\left[\bm S\cdot(\bm\pi\times(\nabla\times\bm B)-(\nabla\times\bm B)\times\bm\pi) \right.\\ \left.
+\bm S\cdot\nabla(\bm S\cdot(\nabla\times\bm B))+\nabla(\bm S\cdot(\nabla\times\bm B))\cdot\bm S\right].
\end{array} \label{commret} \end{equation}

We may use the Maxwell equation
\begin{equation}
\nabla\times \bm B=4\pi\bm j+\frac{\partial \bm E}{\partial t}. \label{Maxwell} \end{equation}
In the absence of time-dependent electric field $\bm E(t)$, a nonzero value of commutator (\ref{commret}) can be caused only by a contact interaction of particle with the external current $\bm j$.
Importantly, such a current either does not appear or can be neglected in accelerators and storage rings.

The exact FW transformation in a stationary \emph{nonuniform}
magnetic field can be fulfilled only for spin-1 particles with $g=2$, as well as for spin-1/2 particles \cite{C}. Equations (\ref{commret}) and (\ref{Maxwell}) define the additional condition of the exact transformation for spin-1 particles ($\bm j=0$). Since this condition is sufficient but is not necessary, formula (\ref{commret}) does not prove an absence of the exact FW transformation in the case of $\bm j\neq0$.

When $\bm j=0$, the exact FW Hamiltonian is given by 
\begin{equation}
{\cal H}_{FW}=\rho_3\sqrt{m^2+\bm\pi^2-2e\bm S\cdot\bm B}.
\label{eqtnHFW}
\end{equation}

It is important that the term proportional to the spin operator can be exactly extracted from the square root.
The extraction can be made when the magnetic field is uniform. A similar extraction has been performed for a spin-1/2 particle with an AMM \cite{JETP1998E}. An important specific property of spin-1 particles is an appearance of a term quadratic in spin. 
%
When the magnetic field is uniform, $\bm B=B\bm e_z$, $[\bm\pi^2,\bm B]=0$, and the FW Hamiltonian can be presented in the form
\begin{equation}
{\cal H}_{FW}
=\rho_3\epsilon'\left(1+a_1S_z+a_2S_z^2\right),~~~\epsilon'=\sqrt{m^2+\bm\pi^2}.
\label{eqtnR}
\end{equation} 
Since $S_z^3=S_z,~S_z^4=S_z^2$, squaring Eq. (\ref{eqtnR}) allows us to
%
obtain
\begin{equation}\begin{array}{c}
a_1=-\frac{\sqrt{1+b}-\sqrt{1-b}}{2}, \\a_2=-\left(1-\frac{\sqrt{1+b}+\sqrt{1-b}}{2}\right), ~~~ b=\frac{2eB}{{\epsilon'}^2}.
\end{array}\label{aoneatwo}
\end{equation} Eigenvalues of the $S_z$ operator are equal to $s_z=0,\pm1$.


Evidently, energy levels for three spin projections ($s_z=0,\pm1$) are not equidistant. However, this circumstance does not lead to any observable effects. Since the operator $\bm\pi^2$ commutes with the Hamiltonian, its eigenvalues are defined by the well-known Landau formula:
\begin{equation}\begin{array}{c} \bm\pi^2_n=|e|B(2n+1), ~~~ n=0,1,2,\dots
\end{array}\label{eigenpi}\end{equation}
As a result, the energy levels are degenerate. The degree of degeneracy of energy levels with $n-es_z/|e|\ge1$ is equal to three.

Dirac particles possess a similar property but their energy levels (except for the lower level) in a uniform magnetic field are twice degenerate.

Spin-tensor effects described in Refs. \cite{Bar3,PRC2008,PRC2009} \emph{do not appear} for the particle with the normal magnetic moment. When the magnetic field is uniform,
the conservation of projections of the polarization operator onto the directions
\begin{equation}
\bm B,~~~\bm \pi\times\bm B, ~~~ \bm B\times(\bm \pi\times\bm B)
\label{three}
\end{equation} and also onto the
direction
$\bm \pi$ takes place \cite{EPJC}. Therefore, we can choose the directions (\ref{three}) as the three basic orthogonal directions.

We can also prove the conservation of projection of the polarization operator onto the binormal direction of the Frenet-Serret coordinate system, $\bm{\mathfrak{B}}=\bm \pi\times(\bm \pi\times\bm B)$ \cite{footnote}. Since $[\bm \pi,(\bm \pi\cdot\bm B)]=0$ and $\bm\Pi\cdot\bm \pi=\Pi_\pi|\bm \pi|$, it can be easily proven that 
$$\bm\Pi\cdot\bm{\mathfrak{B}}=\Pi_\pi|\bm \pi|(\bm \pi\cdot\bm B)-
\Pi_zB\bm \pi^2.$$ All operators in the right-hand side of this equation commute with the FW Hamiltonian. On the other hand,
$\bm\Pi\cdot\bm{\mathfrak{B}}=\Pi_\mathfrak{B}|\bm{\mathfrak{B}}|.$
The operator $|\bm{\mathfrak{B}}|=\sqrt{\bm \pi^4B^2-\bm \pi^2(\bm \pi\cdot\bm B)^2}$ and therefore the binormal projection of the polarization operator commute with the Hamiltonian (\ref{eqtnHFW}) and are conserved. Thus, the projections of the polarization operator onto the three directions of the
Frenet-Serret coordinate system,
$\bm \pi,~\bm \pi\times\bm B,$ and $\bm \pi\times(\bm \pi\times\bm B)$, remain unchanged.

Since the projections of the polarization operator onto the directions (\ref{three}) commute with the FW Hamiltonian, all their bilinear combinations like
$\Pi_\pi^2,~\Pi_z\Pi_{\pi B}+\Pi_{\pi B}\Pi_z$ also commute with this Hamiltonian. This proves the conservation of the both vector and tensor polarizations in the coordinate system (\ref{three}) and the Frenet-Serret one.

Certainly, an addition of terms quadratic in spin to the Hamiltonian (\ref{eqtnHFW}) results in an appearance of the effects described in Refs. \cite{Bar3,PRC2008,PRC2009}.

\section{Foldy-Wouthuysen Hamiltonian for a spin-1 particle with an anomalous magnetic moment}

When a particle possesses an AMM, the spin dynamics becomes much more complicated as compared with the case described in the precedent section. Main effects taking place in a magnetic field have been phenologically described in Refs. \cite{Bar3,PRC2008}. In framework of the relativistic quantum mechanics, the spin dynamics of a \emph{pointlike} particle can be calculated with a high precision. For this purpose, the initial Hamiltonian in the Sakata-Taketani representation (\ref{eqXVI}) should be transformed to the FW representation. In Ref. \cite{TMPFW}, this transformation has been performed for a particle with an AMM moving in the plane orthogonal to the field direction. When the magnetic field is uniform, the operator $\pi_z=p_z$ commutes with the Hamiltonian (\ref{eqXVI}). The FW transformation of the operator $\pi_z$ does not change its form. Therefore, this operator also commutes with the FW Hamiltonian and has eigenvalues ${\cal P}_z\!=\!{\rm const}$. Consequently, a consideration of the particular case ${\cal P}_z\!=\!0$ is quite reasonable \cite{TMPFW}. In this case, the coordinate system (\ref{three}) and the Frenet-Serret one are equivalent. In the classical limit, they are also equivalent to the cylindrical coordinate system \cite{EPJC}.

The FW Hamiltonian calculated in Ref. \cite{TMPFW} up to terms of the order of $|e|^3B^3/m^5$ is given by
\begin{equation}\begin{array}{c} {\cal H}_{FW}=
\rho_3\epsilon-\rho_3\frac{e(g-2)}{2m} \bm
S\cdot\bm B\\+\rho_3\frac{e^2(g-1)(g-2)}{16m^3}
\Biggl\{\frac{1}{\epsilon(\epsilon+m)},\biggl(B^2(\bm S\cdot\bm
\pi)^2\\-[\bm S\cdot(\bm \pi\times\bm B)]^2-e(g-1)B^2(\bm
S\cdot\bm B)\biggr)\Biggr\},\\\epsilon=\sqrt{m^2+\bm
\pi^{2}-2e\bm{S}\cdot\bm B-\frac{e^2g(g-2)}{4m^2}(\bm
S\cdot\bm B)^2}.
\end{array}\label{eqprf}
\end{equation}

The terms proportional to $(\bm S\cdot\bm
B)^2$ and $[\bm S\cdot(\bm \pi\times\bm B)]^2$ define the tensor magnetic and electric polarizabilities of the moving particle, respectively \cite{PRDspin1}. The term proportional to $(\bm S\cdot\bm
\pi)^2$ can be transformed with the equality \cite{TMPFW}
$$B^2(\bm S\cdot\bm
\pi)^2+[\bm S\cdot(\bm \pi\times\bm B)]^2+\bm \pi^2(\bm S\cdot\bm
B)^2=2(\bm \pi\times\bm B)^2.$$
The term in the right-hand side of this equality characterizes the scalar electric polarizability of the moving particle.

Equation (\ref{eqprf}) shows that spin-dependent effects do not vanish for ultrarelativistic particles. This is a common feature of the high-energy spin physics (see Ref. \cite{BarWorldSc}).

Since $[S_z^2,S_x^2]=[S_z^2,S_y^2]=0$, the operator $S_z^2$ commutes with the Hamiltonian and has definite eigenvalues 0 and 1. As a contrary, the
operator $S_z$ does not commute with the Hamiltonian. This noncommutativity leads to mixing the states with $s_z=\pm1$, while the state with $s_z=0$ is pure.

It is convenient to use the coordinate system (\ref{three}). Since ${\cal P}_z\!=\!0$, the directions $\bm B\times(\bm \pi\times\bm B)$ and $\bm \pi$ are parallel and $S_{B\pi B}=S_{\pi}$. While all terms proportional to $B^3$ can be properly calculated with the use of Eqs. (\ref{eqtnR}) and (\ref{aoneatwo}), they are negligible and may be omitted. This circumstance allows us to replace $\epsilon$ with $\epsilon'$ in terms proportional to $B^2$ and to present Eq. (\ref{eqprf}) in the form
\begin{equation}\begin{array}{c} {\cal H}_{FW}=\rho_3\sqrt{m^2+\bm
\pi^{2}-2eS_zB}\\-\rho_3\frac{e(g-2)}{2m}S_zB -\rho_3\frac{e^2g(g-2)}{8m^2\epsilon'}
S_z^2B^2\\-\rho_3\frac{e^2(g-1)(g-2)}{16m^3}\left\{\frac{\epsilon'-m}{\epsilon'},
(S_{\pi B}^2-S_{\pi}^2)\right\}B^2.
\end{array}\label{wqpqf}
\end{equation}

Since a noncommutativity of the operator $\bm\pi^2$ with the operators $S_{\pi B}^2$ and $S_{\pi}^2$ can be neglected, the eigenvalues of $\bm\pi^2$ are defined by the Landau formula (\ref{eigenpi}).

\section{Stationary polarization of particles in a uniform magnetic field}\label{New}

Determination of a particle polarization in stationary states demonstrates a deep difference between polarization properties of spin-1/2 and spin-1 particles in a uniform magnetic field.
For spin-1/2 particles, there are two stationary polarization states with $s_z=\pm1/2$. Spin-1 particles possess fundamentally different properties.

To determine these properties, we should derive eigenvalues and eigenfunctions of Hamiltonian (\ref{wqpqf}) and the spin operators. We use the conventional definition of spin matrices:
\begin{equation}
\begin{array}{c}
S_x=\frac{1}{\sqrt{2}}\left(\begin{array}{ccc} 0 &
 1 & 0\\ 1 & 0 & 1\\ 0 & 1 & 0
 \end{array}\right), ~~~ S_y=\frac{i}{\sqrt{2}}\left(\begin{array}{ccc} 0 &
 -1 & 0\\ 1 & 0 & -1\\ 0 & 1 & 0
 \end{array}\right), \\ S_z=\left(\begin{array}{ccc} 1 &
 0 & 0\\ 0 & 0 & 0\\ 0 & 0 & -1
 \end{array}\right).            
 \end{array}\label{eqsm}\end{equation}

We need to express the spin projections onto the transversal ($\bm \pi\times\bm B$) and longitudinal ($\bm \pi$) directions in terms of the matrices $S_x$ and $S_y$. One can
present the spin matrices as follows,
\begin{equation} \begin{array}{c}
S_{\pi B}=\sum_{i=x,y,z}\frac{(\bm \pi\times\bm B)_i}{|\bm \pi\times\bm B|}S_i,~~~ S_\pi=\sum_{i=x,y,z}\frac{\pi_i}{|\bm \pi|}S_i,
 \end{array}\label{eqsmoll}\end{equation} where the noncommutativity of the operators $\pi_i$ and $|\bm \pi|$ is neglected. However, calculations with Eq. (\ref{eqsmoll}) are very cumbersome.
The replacement $S_{\pi B}\rightarrow S_x,~S_\pi\rightarrow S_y$ is much more convenient and results in
\begin{equation}
\begin{array}{c}
S_{\pi B}=\frac{1}{\sqrt{2}}\left(\begin{array}{ccc} 0 &
 1 & 0\\ 1 & 0 & 1\\ 0 & 1 & 0
 \end{array}\right), ~~ S_\pi=\frac{i}{\sqrt{2}}\left(\begin{array}{ccc} 0 &
 -1 & 0\\ 1 & 0 & -1\\ 0 & 1 & 0
 \end{array}\right).
 \end{array}\label{eqsmo}\end{equation}
However, this replacement is a nontrivial procedure which requires a change of the Hamiltonian. The directions $\bm \pi$ and $\bm \pi\times\bm B$ rotate relative to the Cartesian coordinate axes with the angular velocity equal to the instantaneous angular velocity of orbital revolution of the particle, $\bm\omega$. The instantaneous angular velocity of spin rotation in the frame (\ref{three}) is therefore defined by (cf. Ref. \cite{CPRSTAB}) $\bm\omega_a=\bm\Omega-\bm\omega$,
where $\bm\Omega$ is the instantaneous angular velocity of spin rotation relative to the Cartesian coordinate system.

For the particle with the normal magnetic moment, $\bm\omega$ is exactly equal to $\bm\Omega$ and Hamiltonian (\ref{wqpqf}) contains only the first term [cf. Eq. (\ref{eqtnHFW})]. The elimination of this term restores the right description of spin dynamics $(\bm\omega_a=0)$ when the definition (\ref{eqsmo}) of the spin matrices is used.

In the general case, the situation remains unchanged. The contribution of the above mentioned term into the total angular velocity of spin rotation is always equal to $\bm\omega$. Thus, the use of the definition (\ref{eqsmo}) should be accompanied by the elimination of the first term in Hamiltonian (\ref{wqpqf}). Instead of the elimination, this term may be replaced for the unimportant spin-independent term ${\cal H}_0$ commuting with the operator $\bm\pi^2$. With this replacement, the spin components $S_{\pi B}$ and $S_\pi$ remain unchanged for the particle with the normal magnetic moment. 

The same argumentation is valid for spin-tensor effects. The commutativity of the operators $S_{\pi B}$ and $S_\pi$ with Hamiltonians (\ref{eqtnHFW}) and (\ref{eqtnR}) is a result of joint action of spin-vector and spin-tensor interactions. The use of the definition (\ref{eqsmo}) should not bring spin-tensor effects caused by the noncommutativity of the matrix $S_z^2$ with the matrices $S_x$ and $S_y$ (when they define the operators $S_{\pi B}$ and $S_\pi$). The elimination of the first term in Hamiltonian (\ref{wqpqf}) perfectly solves this problem.

We can mention that three lower components of the six-component wave function correspond to the lower Dirac spinor. In the FW representation, they are always equal to zero for positive-energy states. Therefore, we can omit these components and present the FW Hamiltonian in the form
\begin{equation}\begin{array}{c} {\cal H}_{FW}={\cal H}_0+\omega_0
S_z+\zeta S_z^2+\kappa(S_{\pi B}^2-S_\pi^2),
\end{array}\label{wqpqb}
\end{equation} where the definition (\ref{eqsmo}) is used and
\begin{equation}\begin{array}{c} \omega_0=-\frac{e(g-2)}{2m}B,~~~
\zeta=-\frac{e^2g(g-2)}{8m^2\epsilon'}B^2,\\ \kappa=-\frac{e^2(g-1)(g-2)(\epsilon'-m)}{8m^3\epsilon'}B^2,\\
S_z^2=\left(\begin{array}{ccc} 1 &
 0 & 0\\ 0 & 0 & 0\\ 0 & 0 & 1
 \end{array}\right), ~~ S_{\pi B}^2-S_\pi^2=\left(\begin{array}{ccc} 0 &
 0 & 1 \\ 0 & 0 & 0\\ 1 & 0 & 0
 \end{array}\right).
\end{array}\label{vqpqn}\end{equation}
While the operators $\bm\pi^2$ and $\epsilon'$ do not commute with the original operators $S_\pi^2$ and $S_{\pi B}^2$, they commute with the matrices $S_x^2$ and $S_{y}^2$. Therefore, the anticommutator contained in Eq. (\ref{wqpqf}) is omitted.

The term $\omega_0 S_z$ in Eq. (\ref{wqpqb}) violates the degeneracy of energy levels and defines the spin rotation with the angular frequency $\omega_0$. The next two terms cause spin-tensor effects. The matrix form of Eq. (\ref{wqpqb}) is given by
\begin{equation}\begin{array}{c} {\cal H}_{FW}=\left(\begin{array}{ccc} {\cal H}_0+\omega_0+\zeta &
 0 & \kappa \\ 0 & {\cal H}_0 & 0\\ \kappa & 0 & {\cal H}_0-\omega_0+\zeta
 \end{array}\right).
\end{array}\label{wqpmf}
\end{equation}

The eigenvalues and eigenvectors of the FW Hamiltonian defining the particle energy are equal to
\begin{equation}\begin{array}{c} {\cal H}_{FW}\Psi_i=E_i\Psi_i~~~(i=+1,0,-1),\\
E_{\pm1}={\cal H}_0\pm\omega_0\sqrt{1+\beta^2}+\zeta,~~~E_{0}={\cal H}_0,  \\
\Psi_{+1}=\exp{(i\chi_{+1})}\left(\begin{array}{c} \frac{1\!+\!\sqrt{1\!+\!\beta^2}}{Z} \\
 0 \\ \frac{\beta}{Z}
 \end{array}\right), \\ \Psi_{0}=\exp{(i\chi_{0})}\left(\begin{array}{c} 0 \\
 1 \\ 0
 \end{array}\right),\\ \Psi_{-1}=\exp{(i\chi_{-1})}\left(\begin{array}{c} -\frac{\beta}{Z} \\
 0 \\ \frac{1\!+\!\sqrt{1\!+\!\beta^2}}{Z}
 \end{array}\right),\\
\beta=\frac{\kappa}{\omega_0}=\frac{e(g-1)(\epsilon'-m)}{4m^2\epsilon'}B, \\ Z=\sqrt{2\sqrt{1+\beta^2}\left(1+\sqrt{1+\beta^2}\right)}.
\end{array}\label{eigenvv}\end{equation}
Appropriate eigenvalues of the operator $\epsilon'$ deduced from Eq. (\ref{eigenpi}) should be substituted into Eq. (\ref{eigenvv}). The three last terms in Hamiltonian (\ref{wqpqb}) violate the 
degeneracy of energy levels.

Particle polarization in the stationary states is given by
\begin{equation}\begin{array}{c} <\pm1|S_{z}|\pm1>=\pm Y,~~~<\pm1|S_{\pi B}^2|\pm1>=\frac{1\pm\beta Y}{2},\\
<\pm1|S_{\pi}^2|\pm1>=\frac{1\mp\beta Y}{2},~~~<\pm1|S_{z}^2|\pm1>=1,\\
<0|S_{z}|0>=0,~~~<0|S_{\pi B}^2|0>=<0|S_{\pi}^2|0>=1,\\
~~~<0|S_{z}^2|0>=0,~~~ Y=\frac{1}{\sqrt{1+\beta^2}}.
\end{array}\label{eigenpp}\end{equation}
In any stationary state,
\begin{equation}\begin{array}{c} <S_{\pi B}>=<S_{\pi}>=0, \\  \left\langle\{ S_{\pi B},S_{\pi}\}\right\rangle=\left\langle\{S_{\pi B},S_{z}\}\right\rangle=\left\langle\{S_{\pi},S_{z}\}\right\rangle=0.
\end{array}\label{additio}\end{equation}

Equations (\ref{eigenpp}) and (\ref{additio}) demonstrate new unexpected properties of stationary polarization of pointlike spin-1 particles in a uniform magnetic field. The spin projection onto the magnetic field direction is \emph{not} integer. The tensor polarization is also nontrivial and differs for the transversal ($\bm \pi\times\bm B$) and longitudinal ($\bm \pi$) directions. The effects are small and are of the orders of $\beta$ and $\beta^2$ for tensor and vector effects, respectively. However, they are fundamentally important because they show a deep difference between polarization properties of spin-1/2 and spin-1 particles. In particular, polarization properties of spin-1 particles cannot be exhaustively described by the rotation group $SO(3)$.

For a particle at rest, $\epsilon'=m,~\beta=0$. New spin properties defined by Eq. (\ref{eigenpp}) disappear. In this case, spin-tensor effects are caused by the term $\zeta S_z^2$ in Eq. (\ref{wqpqb}). Amazingly, these effects exist not only for extended objects like nuclei \cite{Bar3} but also for pointlike particles. The tensor interaction of the spin produces the spin rotation with two frequencies instead of
one, beating with a frequency proportional to the tensor magnetic polarizability, and causes
transitions between vector and tensor polarizations \cite{Bar3}. In particular, an initially tensor-polarized beam 
acquires a final horizontal vector polarization \cite{PRC2008}.

\section{Discussion and summary}

The results obtained allow us to ascertain new properties of a spin-1 particle in a magnetic field.
The exact FW Hamiltonian has been found for a particle with a normal magnetic moment ($g=2$) in a \emph{nonuniform} magnetic field. For such a particle, the vector and tensor polarizations defined relative to the momentum direction are conserved in a \emph{uniform} magnetic field.
The energy spectrum 
is rather simple and the degree of degeneracy of almost all energy levels 
is equal to three. The spin projections onto the direction of the uniform magnetic field are integer. The projections of the polarization operator not only onto this direction but also onto the
directions
$\bm \pi, ~ \bm \pi\times\bm B$, $\bm B\times(\bm \pi\times\bm B)$ \cite{EPJC}, and the
directions of the Frenet-Serret coordinate system are conserved.

The situation is different for a pointlike spin-1 particle with an AMM ($W$ boson). Its stationary polarization in a uniform magnetic field possesses unexpected properties. If the particle moves, projections of its spin onto the
magnetic field direction are not integer and the tensor polarization is asymmetric in the horizontal plane. Moving extended objects (nuclei) with spin 1 and higher spins also possess these properties. However, the existence of such properties was not established in previous investigations. Since the tensor electric and magnetic
polarizabilities of nuclei are much greater than those of the W boson \cite{Spinunit}, the effects described should be searched just for nuclei. Because mixing of states with $s=1$ and $s=0$ should be negligible, experimental investigations cannot be made with atoms. Since the energy splitting is small, a measurement of observable anomalies in the spin motion (see Refs. \cite{PRC2008,PRC2009}) is the best way to search for the above effects. These effects are fundamentally important because they establish new spin properties.

To calculate the spin dynamics, previously obtained results can be used. In particular, the equality $S_{\pi B}^2+S_\pi^2+S_z^2=2$ allows one to apply equations of spin dynamics derived in Ref. \cite{PRC2009} after the substitutions ${\cal A}\rightarrow\kappa,~({\cal B}-{\cal A})\rightarrow\zeta,~\omega'\rightarrow\omega_0\sqrt{1+\beta^2}$.

For a particle at rest, the new spin properties found in the previous section disappear. Importantly, spin-tensor
effects described is Refs. \cite{Bar3,PRC2008,PRC2009} and caused by the tensor magnetic
polarizability occur not only for nuclei but also for pointlike particles.

\section*{Acknowledgements}
The author acknowledges the support by the Belarusian Republican Foundation for Fundamental Research
(Grant No. $\Phi$12D-002).



\begin{thebibliography}{}
\bibitem{footnote}
In the considered case, the tangent and normal unit vectors of the Frenet-Serret coordinate system are collinear to the particle velocity (parallel to $\bm\pi$) and the Lorentz force (parallel to $\bm\pi\times\bm B$), respectively.
\bibitem{PRDspin1}
The tensor magnetic and electric polarizabilities of a particle at rest have been recently calculated in
Ref. \cite{Spinunit}.
\end{thebibliography}
\end{document}